# The Symmetries of the $\pi$-metric

Marcelo Muniz Silva Alves * Luciano Panek†

**Abstract**

Let $V$ be an $n$-dimensional vector space over a finite field $\mathbb{F}_q$. We consider on $V$ the $\pi$-metric $d_\pi$ recently introduced by K. Feng, L. Xu and F. J. Hickernell. In this paper we give a complete description of the group of symmetries of the metric space $(V, d_\pi)$.

## 1 Introduction

Let $\mathbb{F}_q^n$ be the finite field of $q$ elements. Let $n$ be a positive integer and let $\pi$ be a partition of $n$, i.e., $\pi$ is a non-empty sequence $(k_1, k_2, \ldots, k_m)$ where each $k_i$ is a positive integer, $n = k_1 + \ldots + k_m$, $m \geq 1$ and $k_1 \geq k_2 \geq \ldots \geq k_m \geq 1$. Feng, Xu and Hickernell introduced in [1] a metric associated to the partition $\pi$, the $\pi$-*metric*, similar to the classic Hamming metric. The partition $\pi$ induces a direct sum decomposition

$$\mathbb{F}_q^n = \mathbb{F}_q^{k_1} \oplus \mathbb{F}_q^{k_2} \oplus \ldots \oplus \mathbb{F}_q^{k_m}$$

where every vector $v$ of $\mathbb{F}_q^n$ is written as

$$v = (v_1, v_2, \ldots, v_m),$$

with $v_i \in \mathbb{F}_q^{k_i}$. The $\pi$-*weight* of $v = (v_1, v_2, \ldots, v_m) \in \mathbb{F}_q^n$ is

$$\omega_\pi(v) = |\{i : 1 \leq i \leq m, v_i \neq 0\}|$$

and the $\pi$-distance between $u$ and $v$ is given by $d_\pi(u, v) = \omega_\pi(u - v)$. Note that when $k_i = 1$ for all $i$ (and $m = n$) the $\pi$-weight is the Hamming weight over $\mathbb{F}_q^n$. It is clear that

$$d_\pi(u, v) = |\{i : 1 \leq i \leq m, u_i \neq v_i\}|.$$

A $\mathbb{F}_q$-linear code in $(\mathbb{F}_q^n, d_\pi)$ is called a *linear error-block code.* The linear symmetries (automorphisms) of $\mathbb{F}_q^n$ with respect to a $\pi$-metric have already been determined in [2]. In this paper we describe all symmetries of $(\mathbb{F}_q^n, d_\pi)$; in order to do that, we describe two subgroups $\mathcal{M}$ and $S_\pi$ of symmetries and prove that the full symmetry group is their semi-direct product. We also reobtain the automorphism group as a particular case.

From this point on we write simply $d$ and $\omega$ instead of $d_\pi$ and $\omega_\pi$.

*Departamento de Matemática, Centro Politécnico, UFPR, Caixa Postal 019081, 81531-990, Curitiba, PR. E-mail: marcelomsa@ufpr.br

†Centro de Engenharias e Ciências, UNIOESTE, 85870-650, Foz do Iguaçu, PR. E-mail: lucpanek@gmail.com

## 2 The Symmetry Group

A *symmetry* of $\left(\mathbb{F}_q^n, d\right)$ is a bijection $\varphi : \mathbb{F}_q^n \to \mathbb{F}_q^n$ that preserves distance, i.e.,

$$d\left(\varphi\left(u\right), \varphi\left(v\right)\right) = d\left(u, v\right)$$

for every $u, v \in \mathbb{F}_q^n$. An *automorphism* is a linear symmetry. We will denote the group of symmetries of $\left(\mathbb{F}_q^n, d\right)$ by $Symm\left(\mathbb{F}_q^n, d\right)$.

Let $\mathcal{M}$ be the group of mappings $T : \mathbb{F}_q^n \to \mathbb{F}_q^n$, where

$$T\left(\left(v_1, v_2, \ldots, v_m\right)\right) = \left(T_1\left(v_1\right), \ldots, T_m\left(v_m\right)\right)$$

and every "coordinate function" $T_i$ is a bijection from $\mathbb{F}_q^{k_i}$ onto itself. Since each $T_i$ is a bijection, $T_i\left(v_i\right) \neq T_i\left(u_i\right)$ if and only if $u_i \neq v_i$. It follows that

$$d\left(T\left(u\right), T\left(v\right)\right) = \sum_{i=1}^{m} \delta\left(T_i\left(u_i\right), T_i\left(v_i\right)\right) = \sum_{i=1}^{m} \delta\left(u_i, v_i\right) = d\left(u, v\right),$$

so that every $T \in \mathcal{M}$ is a symmetry of $\left(\mathbb{F}_q^n, d\right)$.

Let $B(\mathbb{F}_q^{k_i})$ be the group of bijections from $\mathbb{F}_q^{k_i}$ onto itself. Clearly, if

$$M_i = \left\{\left(id, \ldots, id, T_i, id, \ldots, id\right) \in \mathcal{M} : T_i \in B(\mathbb{F}_q^{k_i})\right\}$$

then $\mathcal{M} = M_1 \times \ldots \times M_m$. Besides, $M_i$ is isomorphic to $B(\mathbb{F}_q^{k_i})$, which is in turn isomorphic to $S_{q^{k_i}}$, the permutation group on a set of $q^{k_i}$ symbols; hence, we have proved the following:

**Proposition 1** *Let $\mathcal{M}$ be the group of mappings $T : \mathbb{F}_q^n \to \mathbb{F}_q^n$ defined above. Then*

*(i) $\mathcal{M}$ is a subgroup of $\left(\mathbb{F}_q^n, d\right)$.*

*(ii) $\mathcal{M}$ is isomorphic to the direct product $\prod_{i=1}^{m} S_{q^{k_i}}$ .*

Let $S_m$ be the permutation group of $\{1, 2, \ldots, m\}$. We will call a permutation $\sigma \in S_m$ *admissible* if $\sigma\left(i\right) = j$ implies that $k_i = k_j$. Clearly, the set $S_\pi$ of all admissible permutations is a subgroup of $S_m$.

**Proposition 2** *[2] $S_\pi$ acts as a group of linear symmetries in $V$.*

Besides being linear, these symmetries satisfy an important property: if

$$V_i = \left\{\left(0, \ldots, 0, v_i, 0, \ldots, 0\right) \in \mathbb{F}_q^n : v_i \in \mathbb{F}_q^{k_i}\right\},$$

then $\sigma\left(V_i\right) = V_{\sigma(i)}$. As we see in the following, even non-linear symmetries satisfy an analogous property.

**Lemma 1** *Let $F$ be a symmetry of $\left(\mathbb{F}_q^n, d\right)$ and let $v \in \mathbb{F}_q^n$. For each $i \in \{1, 2, \ldots, m\}$, there exists $j \in \{1, 2, \ldots, m\}$ such that $k_i = k_j$ and*

$$F\left(v + V_i\right) = F\left(v\right) + V_j.$$

**Proof.** Let $v_i \in V_i$, $v_i \neq 0$. Since $d\left(F\left(v+v_i\right), F\left(v\right)\right) = d\left(v+v_i, v\right) = 1$, $F\left(v+v_i\right) - F\left(v\right)$ is a vector of $\pi$-weight 1. But for any vector $u$ in $\mathbb{F}_q^m$, $\omega\left(u\right) = 1$ iff $u \in V_j$ for some $j$. Hence there is an index $j$ such that $F\left(v+v_i\right) = F\left(v\right) + v_j$, with $v_j \in V_j$, $v_j \neq 0$.

If $v_i' \in V_i$, $v_i' \neq v_i$ and $v_i' \neq 0$, then $F\left(v+v_i'\right) = F\left(v\right) + v_k$ for some $v_k \in V_k$ (with $v_k \neq 0$), but also $d\left(F\left(v+v_i\right), F\left(v+v_i'\right)\right) = d\left(v+v_i, v+v_i'\right) = 1$. If $k \neq j$ then $d\left(F\left(v\right)+v_j, F\left(v\right)+v_k\right) = 2$; hence $k = j$ and $F\left(v+v_i'\right) = v + v_j'$, with $v_j' \neq v_j$. This proves that $F\left(v+V_i\right) \subseteq F\left(v\right) + V_j$.

Now apply the same reasoning to $F^{-1}$: if $v_i \neq 0$ and $F\left(v+v_i\right) = F\left(v\right) + v_j$, then $F^{-1}\left(F\left(v\right)+v_j\right) \in v+V_i$ and therefore $F^{-1}\left(F\left(v\right)+v_j\right) \subseteq v+V_i$. It follows that $F\left(v+V_i\right) = F\left(v\right) + V_j$.

Finally, $k_i = \dim\left(V_i\right) = \dim\left(V_j\right) = k_j$ because $F$ is bijective. □

This result implies, in particular, that for each $i$ there corresponds a $j$ such that $F\left(V_i\right) = F\left(0\right) + V_j$ and that $k_i = \dim\left(V_i\right) = \dim\left(V_j\right) = k_j$. This suggests the following definition.

**Definition 1** *Let $F$ be a symmetry of $\left(\mathbb{F}_q^n, d\right)$. The admissible permutation $\sigma_F$ induced by $F$ is given by*

$$\sigma_F\left(i\right) = j \text{ if and only if } F\left(V_i\right) = F\left(0\right) + V_j.$$

**Theorem 1** *Every symmetry $F$ is a product $\sigma T$, where $\sigma \in S_\pi$ and $T \in \mathcal{M}$. This decomposition is unique.*

**Proof.** Suppose that $F\left(0\right) = 0$; then $F = \left(\sigma_F\right)\left(\sigma_F^{-1} F\right)$, where $\sigma_F$ is the admissible permutation defined by $F$. Clearly, $\left(\sigma^{-1}F\right)\left(V_i\right) = V_i$ for each $i$, and therefore this mapping is in $\mathcal{M}$.

If $F\left(0\right) = v_0 \neq 0$, let $S$ be the translation $S(v) = v - v_0$; then $S \in \mathcal{M}$, $SF(0) = 0$ and, therefore, $SF = \sigma_{SF} T$, where $T \in \mathcal{M}$. It follows that $F$ can be written as $F = \sigma_{SF}(TS^{-1})$.

If follows that every symmetry $F$ is a product $\sigma T$; since $S_\pi \cap \mathcal{M} = \{id\}$, this decomposition is unique. □

**Theorem 2** *(i) $Symm\left(\mathbb{F}_q^n, d\right) \cong S_\pi \ltimes \mathcal{M}$, where the semi-direct structure is induced by the action of $S_\pi$ on $\mathcal{M}$ by conjugation.*

*(ii) $Symm\left(\mathbb{F}_q^n, d\right) \cong S_\pi \ltimes \prod_{i=1}^m S_{q^{k_i}}$.*

**Proof.** We will show that the group $\mathcal{M}$ is normal in $Symm\left(\mathbb{F}_q^n, d\right)$. Since the last result shows that $Symm\left(\mathbb{F}_q^n, d\right) = S_\pi \mathcal{M}$, it is enough to check that if $T = \left(T_1, T_2, \ldots, T_m\right) \in \mathcal{M}$ and $\sigma \in S_\pi$ then $\sigma^{-1} T \sigma$ is also in $\mathcal{M}$. Applying $\sigma^{-1} T \sigma$ to a vector we conclude that

$$\sigma^{-1} T \sigma = \left(T_{\sigma(1)}, \ldots, T_{\sigma(m)}\right)$$

and hence that $\mathcal{M}$ is a normal subgroup of $Symm\left(\mathbb{F}_q^n, d\right)$. Since $S_\pi \cap \mathcal{M} = \{id\}$, this shows that the symmetry group is isomorphic to the semi-direct product $S_\pi \ltimes \mathcal{M}$, where multiplication is given by

$$\left(\sigma, T\right)\left(\varphi, S\right) = \left(\sigma\varphi, \left(\varphi^{-1} T \varphi\right) S\right).$$

The fact that $\mathcal{M} \cong \prod_{i=1}^m S_{q^{k_i}}$ yields the second assertion. □

**Corollary 1** *Let $\pi = (k_1, k_2, \dots, k_m)$ be a partition of $n$. If $k_1 = \dots = k_{m_1} = l_1, \dots, k_{m_1+\dots+m_{l-1}+1} = \dots = k_n = l_r$ with $l_1 > l_2 > \dots > l_r$, then*

$$\left|Symm\left(\mathbb{F}_q^n, d\right)\right| = \left(\sum_{j=1}^{l} m_j!\right) \cdot \left(\prod_{i=1}^{m} \left(q^{k_i}\right)!\right).$$

**Proof.** In fact, since $Symm\left(\mathbb{F}_q^n, d\right) \cong S_\pi \ltimes \prod_{i=1}^{m} S_{q^{k_i}}$ and $S_\pi \cong \prod_{j=1}^{m} S_{l_j}$,

$$\left|Symm\left(\mathbb{F}_q^n, d\right)\right| = |S_\pi| \cdot \prod_{i=1}^{m} \left|S_{q^{k_i}}\right| = \left(\sum_{j=1}^{l} m_j!\right) \cdot \left(\prod_{i=1}^{m} \left(q^{k_i}\right)!\right).$$

□

When $m = n$, $k_1 = k_2 = \dots = k_n = 1$, the $\pi$-weight is the Hamming weight on $\mathbb{F}_q^n$. In this case each $M_i$ is isomorphic to $S_q$ and every permutation in $S_n$ is admissible; thus we reobtain the symmetry groups of Hamming spaces from our previous calculations.

**Corollary 2** *Let $d_H$ be the Hamming metric over $\mathbb{F}_q^n$. The symmetry group of $\left(\mathbb{F}_q^n, d_H\right)$ is isomorphic to $S_n \ltimes S_q^n$.*

Another proof of this can be found in [3].

# 3 Automorphisms

The group of automorphisms of $\left(\mathbb{F}_q^n, d\right)$ is easily deduced from the results above. Let $F = \sigma T$ be a symmetry. Since $\sigma$ is linear, the linearity of $F$ is a matter of whether $T$ is linear or not. Now, if $T = (T_1, T_2, \dots, T_m)$ is linear, then each component $T_i$ must also be linear; since each $T_i$ is bijective, $T_i$ is in the group $Aut\,(V_i)$ of linear automorphisms of $V_i$. Therefore $T \in \prod_{i=1}^{m} Aut\,(V_i)$. On the other hand, any element of this group is a linear symmetry; hence:

**Theorem 3** *The automorphism group $Aut\left(\mathbb{F}_q^n, d\right)$ of $\left(\mathbb{F}_q^n, d\right)$ is isomorphic to $S_\pi \ltimes \prod_{i=1}^{m} Aut\,(V_i)$.*

**Corollary 3** *Let $\pi = (k_1, k_2, \dots, k_m)$ be a partition of $n$. If*

$$k_1 = \dots = k_{m_1} = l_1, \dots, k_{m_1+\dots+m_{l-1}+1} = \dots = k_n = l_r$$

*with $l_1 > l_2 > \dots > l_r$, then*

$$\left|Aut\left(\mathbb{F}_q^n, d\right)\right| = \left(\sum_{j=1}^{l} m_j!\right) \cdot \left(\prod_{i=1}^{m} \left(q^{k_i} - 1\right)\left(q^{k_i} - q\right) \dots \left(q^{k_i} - q^{k_i - 1}\right)\right).$$

**Proof.** Note initially that there is a bijection from $Aut\,(V_i)$ and the family of all ordered bases of $V_i$: let $(e_1, e_2, \dots, e_{k_i})$ be an ordered basis of $V_i$; if $T \in Aut\,(V_i)$, then $(T(e_1), T(e_2), \dots, T(e_{k_i}))$ is an ordered basis of $V_i$; if $(v_1, v_2, \dots, v_{k_i})$ is an ordered basis of $V_i$ then there exist a unique automorphism $T$ with $T(e_j) = v_j$ for all $j \in \{1, 2, \ldots, k_i\}$. Since the number of ordered basis of $V_i$ equal

$$\left(q^{k_i} - 1\right)\left(q^{k_i} - q\right)\ldots\left(q^{k_i} - q^{k_i - 1}\right)$$

follows that $|Aut(V_i)| = \left(q^{k_i} - 1\right)\left(q^{k_i} - q\right)\ldots\left(q^{k_i} - q^{k_i - 1}\right)$. From above theorem

$$\left|Aut\left(\mathbb{F}_q^n, d\right)\right| = |S_\pi| \cdot \prod_{i=1}^{m} |Aut(V_i)|.$$

Since $|S_\pi| = \sum_{j=1}^{l} m_j!$ the corollary follows. □

Restricting to the Hamming case again, $Aut(V_i) = Aut(\mathbb{F}_q) = \mathbb{F}_q^*$, and $S_\pi = S_n$. Hence:

**Corollary 4** *The automorphism group of* $\left(\mathbb{F}_q^n, d_H\right)$ *is* $S_n \ltimes \left(\mathbb{F}_q^*\right)^n$.